\documentclass[12pt]{article}

\usepackage[utf8]{inputenc}

\usepackage{amsmath, amssymb}

\usepackage{times}
\usepackage[normalem]{ulem}
\useunder{\uline}{\ul}{}

\usepackage[table,xcdraw]{xcolor} 
\usepackage{graphicx}
\graphicspath{ {./images/} }

\usepackage{geometry}
\usepackage[skip=10pt, indent=0pt]{parskip}
\usepackage{hyperref}
\hypersetup{
  colorlinks=true,
  linkcolor=blue,
  filecolor=magenta,      
  urlcolor=cyan,
}

\usepackage{multirow}
\usepackage{tabularx}
\usepackage{booktabs}

\usepackage{float}
\usepackage{lscape}

\usepackage[numbers]{natbib}

\usepackage{listings}
\usepackage{algorithm}
\usepackage{algorithmic}
\usepackage[toc,page]{appendix}

\usepackage{setspace}
\usepackage{tablefootnote}

\usepackage{tikz}
\usetikzlibrary{arrows.meta, positioning, calc, shadings}

\usepackage{subcaption}

\usepackage[nottoc]{tocbibind}

\definecolor{dkgreen}{rgb}{0,0.6,0}
\definecolor{gray}{rgb}{0.5,0.5,0.5}
\definecolor{mauve}{rgb}{0.58,0,0.82}

\makeatletter
\renewenvironment{titlepage}
 {%
  \if@twocolumn
    \@restonecoltrue\onecolumn
  \else
    \@restonecolfalse\newpage
  \fi
  \thispagestyle{plain}%
 }
 {%
  \if@restonecol
    \twocolumn
  \else
    \newpage
  \fi
 }
\makeatother
\begin{document}
\begin{titlepage}
    \begin{center}

        \large
            B.Comp.Dissertation \\
            
        \vspace{7cm}

        \Large
        \textbf{Automated Phishing Detection using URLs and Webpages}
        
        \vspace{2cm}
        
        
        \large
        by\\
        \vspace{0.5cm}
        \large
        WANG HUILIN
            
        \vspace{1.5cm}
            
        \large
        Department of Computer Science \\
        School of Computing \\
        National University of Singapore \\
        
        \vspace{1cm}
        \normalsize
        2023/2024
    \end{center}
\end{titlepage}

\pagenumbering{roman}
\begin{titlepage}
    \phantomsection
    \addcontentsline{toc}{section}{Title}
    \begin{center}
            
        \large
        B.Comp.Dissertation\\
            
        
        \Large
        
        \vspace{3cm}
        
        \Large
        \textbf{Automated Phishing Detection using URLs and Webpages}
        
        \vspace{1.5cm}
        
        \Large
        \Large
        
        \large
        by\\
        \vspace{0.5cm}
        \large
        WANG HUILIN
            
        \vspace{3cm}
            
        \large
        Department of Computer Science \\
        School of Computing \\
        National University of Singapore \\
        
        \vspace{1cm}
        \normalsize
        2023/2024
    \end{center}

    \vspace{4cm}
    \begin{flushleft}
    Project No.: H261170\\
    Advisor: HOOI Kuen Yew Bryan\\
    Deliverables: \\
    \end{flushleft}
\end{titlepage}
\thispagestyle{plain}
\begin{center}
    \Large
    \phantomsection
    \addcontentsline{toc}{section}{Abstract}
    \textbf{Abstract}
\end{center}




Phishing detection is a critical cybersecurity task that involves the identification and neutralization of fraudulent attempts to obtain sensitive information, thereby safeguarding individuals and organizations from data breaches and financial loss. In this project, we address the constraints of traditional reference-based phishing detection by developing an LLM agent framework. This agent harnesses Large Language Models to actively fetch and utilize online information, thus providing a dynamic reference system for more accurate phishing detection. This innovation circumvents the need for a static knowledge base, offering a significant enhancement in adaptability and efficiency for automated security measures.

The project report includes an initial study and problem analysis of existing solutions, which motivated us to develop a new framework. We demonstrate the framework with LLMs simulated as agents (GEPAgent) and detail the techniques required for construction, followed by a complete implementation with a proof-of-concept as well as experiments to evaluate our solution's performance against other similar solutions. The results show that our approach has achieved with accuracy of 0.945, significantly outperforms the existing solution(DynaPhish) by 0.445. Furthermore, we discuss the limitations of our approach and suggest improvements that could make it more effective.

Overall, the proposed framework has the potential to enhance the effectiveness of current reference-based phishing detection approaches and could be adapted for real-world applications.

\vspace{2cm}
Subject Descriptors:

\begin{tabular}{c c}
I.2.4 & Knowledge Representation Formalisms and Methods \\
I.2.7 & Natural Language Processing \\
I.4 & Image Processing and Computer Vision \\
H.3.3 & Information Search and Retrieval\\
\end{tabular}

\vspace{2cm}
\indent 

\pagebreak
\begin{spacing}{2}
\thispagestyle{plain}
\begin{center}
    \Large
    \phantomsection
    \addcontentsline{toc}{section}{Acknowledgements}
    \textbf{Acknowledgements}
\end{center}
I want to express my sincere thanks to Professor HOOI Kuen Yew Bryan, my advisor, for his invaluable mentorship, encouragement, and insightful critiques throughout this project. His profound knowledge and perspectives have been pivotal in refining the idea of this project and enhancing my work.

My appreciation also goes out to Dr. Nay Oo for his important suggestions and thoughtful comments on my work progress. His support was crucial in tackling the challenges encountered especially in the project management, resources and supervision.

A special thanks goes to Mr. Li Yuexin and Mr. Huang Chengyu for their valuable contributions, advice, guidance, and innovative ideas that played a significant role in advancing this project. Their essential supports throughout the project's development was incredibly beneficial.

I appreciate Professor Chang Ee-Chien, my project evaluator, for his thorough review of my work and his feedback, which pointed out areas that needed improvement.

Lastly, I want to acknowledge the support and help from the School of Computing, NUS, which greatly helped with the project's administrative aspects.

I'm truly grateful to all of you for your contributions to making this project a success.
\pagebreak
\end{spacing}

\tableofcontents
\pagenumbering{arabic}
\pagebreak

\section{Introduction}
\subsection{Motivation}
Phishing attacks represent a significant threat, with far-reaching consequences for individuals and organizations alike. For instance, they are implicated in approximately 90\% of data breaches. Moreover, the prevalence of these scams is on the rise; Singapore, for example, witnessed a 2.1-fold increase in reported phishing incidents in 2022. This underscores the urgency of developing more effective countermeasures to protect sensitive information and maintain cybersecurity.

\subsection{Project Objective and Overview}
The primary objective of our project is to enhance reference-based phishing detection by developing a sophisticated LLM agent capable of autonomously identifying phishing threats. This agent mimics human information retrieval and decision-making behaviors, using its robust reasoning abilities to search for and utilize online information as a reference, making decisions that are both explainable and adaptable. This stands in contrast to traditional phishing detection systems that depend on static knowledge bases.

Our investigation began with a thorough examination of the current state of the art in reference-based phishing detection, which included a detailed analysis of prevalent solutions. Following the development of the proof-of-concept, we carried out a series of experiments to assess the efficacy of our model in comparison to those existing solutions.

\section{Literature Review}

\subsection{Phishing Detection}

Phishing is a prevalent form of social engineering attack that exploits human vulnerabilities to exfiltrate sensitive information or assets from victims. Whittaker et al.\cite{whittaker2010large} characterize phishing webpages as those which, ``without permission, purport to act on behalf of a legitimate third party with the intent of misleading viewers into performing actions they would otherwise reserve for a trusted entity."

In their comprehensive study, Khonji et al. \cite{khonji2013phishing} categorize phishing detection methodologies into four distinct groups: Blacklist-based, Heuristic-based, Visual Similarity, and Machine Learning approaches. Each category employs different strategies to identify and mitigate phishing threats, ranging from comparing URLs against known malicious databases to utilizing advanced algorithms that learn to discern phishing patterns. 

With the advent of machine learning, today's phishing detection methodologies have evolved to become more adaptable, often integrating multiple approaches for enhanced accuracy. For instance, Phishpedia \cite{lin2021phishpedia} leverages a reference-based detection system that combines computer vision and deep learning for precise logo identification, targeting specific brands. This is achieved with the fact that phishing webpages usually look ``different" than others, and utilized two deep learning  model to resolve logo recognition as well as brand recognition. Building upon this, PhishIntention \cite{liu2022inferring} incorporates a Credential Requirement Page classifier combine with HTML and phishing screenshot motivated by the fact that phishing attacks frequently aim to harvest sensitive information from victims.

However, brand recognition poses a persistent challenge in reference-based phishing detection, particularly due to the ever-increasing number of new brands emerging daily. The efficacy of brand recognition is heavily contingent upon the comprehensiveness of the protected list; it must be continually updated to include the growing array of brands to maintain its effectiveness. Therefore, Dynaphish \cite{liu2023knowledge} employs the Google Search engine and Google Logo Detector to facilitate the expansion of its knowledge base. It also integrates web interaction capabilities to more effectively identify Credential Requirement Pages, thus improving the system's ability to detect and thwart phishing attempts. 

In addition to logos, URLs are a critical component in phishing detection since adversaries often employ atypical domains for phishing websites. Google Safe Browsing, PhishTank, and OpenPhish, are mainly using dynamic blacklist approaches to detecting phishing webpages by tracking reported URLs \cite{bell2020analysis}. However, such approach is limited when facing phishing website in the wild. Sahingoz et al. \cite{sahingoz2019machine} have developed a real-time detection system that leverages machine learning, extracting features directly from URLs. This system incorporates Word Vectors, NLP-based features, and hybrid features extracted from URLs to enhance the detection process.

\subsection{Generative AI and Autonomous Agents}

Generative modeling artificial intelligence (GAI) operates within an unsupervised or partially supervised machine learning framework. It focuses on creating human-made artifacts using statistical and probabilistic methods. Thanks to advancements in deep learning (DL), generative AI is capable of producing artificial artifacts across various digital mediums, including video, images, graphics, text, and audio. This is achieved by analyzing training examples to learn and replicate their patterns and distribution.\cite{baidoo2023education}

Autonomous agents refer to software applications that operate independently, reacting to changes and occurrences in their surroundings without needing explicit commands from the user or owner. Instead, they act in ways that serve and align with the interests of their owner. \cite{BOSSER20011002} The advance of artificial intelligence, such as existence of GAI, provides the ability of enriching the power and ability of autonomous agents. 

Inspired by The Sims, Park et al. \cite{10.1145/3586183.3606763} proposed a novel concept of generative agents, which are computational entities designed to simulate human-like behavior in interactive applications. These agents are capable of performing everyday activities such as cooking and going to work, as well as engaging in more complex interactions like painting, writing, forming opinions, and initiating conversations. They achieve this by utilizing a large language model that can record their experiences, synthesize memories into higher-level reflections, and dynamically retrieve these memories to plan and adjust their behavior accordingly. Generative agents represent a significant advancement in the simulation of human behavior in interactive systems. 

Another work from Li et al. \cite{shen2024hugginggpt} described a system named HuggingGPT that utilizes large language models (LLMs), like ChatGPT, as a controller to manage various AI models from the machine learning community, specifically Hugging Face, to solve complex AI tasks. The approach hinges on the idea that language serves as a universal interface, enabling the coordination of different models to perform tasks across multiple domains and modalities, including language, vision, and speech. This collaboration enables HuggingGPT to offer impressive results across different tasks, showcasing a step towards achieving artificial general intelligence. 

Some other current works related to generative agents include \cite{yang2023appagent}, \cite{chen2024agent}, \cite{zeng2023agenttuning}, and \cite{yao2022react}. Zhang et al.\cite{yang2023appagent} present an innovative LLM-based multimodal agent framework, tailor-made for smartphone application usage, that employs human-like gestures such as taps and swipes, eliminating the need for system back-end access. Chen et al.\cite{chen2024agent} unveil AgentFLAN, a framework designed to bolster the capabilities of LLMs, enabling them to act effectively as agents. Zeng et al.\cite{zeng2023agenttuning} develop a methodology that leverages a specialized dataset, AgentInstruct, to enhance LLMs for superior performance in complex agent tasks. Lastly, Yao et al.\cite{yao2022react} unveil ReAct, a strategy that augments LLMs' capabilities by generating interlinked reasoning traces and task-specific actions, thereby enhancing the synergy between comprehension and execution. This technique notably improves LLMs' proficiency in forming, monitoring, and revising action plans, as well as in interacting with external information sources.

\section{Investigation of Existing Solution}

Reference-based phishing detection is intrinsically reliant on the breadth and depth of its protect brand list, with a particular emphasis on brand recognition accuracy. Dynaphish, in this case, is chosen for examination to gain a comprehensive understanding of the latest advancements and the impact of external APIs within this domain. The investigation focuses on evaluating Dynaphish's capabilities in terms of knowledge base expansion and its proficiency in brand recognition, which are critical determinants of its overall effectiveness.



\subsection{Reproduction of Dynaphish}


In the study \cite{liu2023knowledge}, Dynaphish is described as comprising several components, including a logo cropper, a representation validation mechanism for the target brand, a web interaction component for CRP classification, a knowledge expansion module designed to augment the existing knowledge base, and a population validation feature for domain verification. Additionally, Dynaphish incorporates two pre-trained CV models: the Phishpedia model \cite{lin2021phishpedia} and the PhishIntention model \cite{lin2021phishpedia}. For the purposes of this research, the component related to CRP classification will be excluded from consideration due to its irrelevance to the scope. The focus will instead be on Dynaphish's capability for knowledge expansion and its performance in scenarios where the knowledge base is either absent or deemed inapplicable, thereby also excluding the knowledge base from the investigation.

\subsection{Trial Run of limited Dynaphish}



Phishing 3k dataset, which consists around 3,310 phishing samples are used to evalutate Dynaphish. This dataset encompasses phishing attempts across approximately 340 distinct brands. Our objective is to assess Dynaphish's performance in terms of its execution time and effectiveness in detecting phishing attempts. Additionally, we aim to explore the system's knowledge expansion capabilities, particularly with the integration of Google search.

To this end, we replaced the original knowledge base with an almost empty knwoledge base with two brands only: Airbnb and ADP. This approach allows for a more targeted analysis of Dynaphish's knowledge expansion module.

Upon processing the dataset with Dynaphish, we observed a recall rate of approximately 40\%. Additionally, the knowledge base underwent significant expansion, growing from the initial two brands to an impressive total of 216 brands.

\begin{table}[htbp]
\centering
\begin{tabular}{|l|c|}
\hline
\multirow{2}{*}{Metric} & Phishing Data \\ \cline{2-2} 
                        & Value \\ \hline
\multirow{2}{*}{Detection} & TP: 1324/3310 ($\approx$ 40\%) \\ \cline{2-2}
                           & FN: 1986/3310 ($\approx$ 60\%) \\ \hline
Knowledge Expansion       & 2 $\rightarrow$ 216 \\ \hline
\end{tabular}
\caption{3k Phish Dataset Statistics}
\end{table}

\subsection{Problem Investigation}



To gain deeper insights into Dynaphish's performance, we conducted a granular analysis by segregating the results according to individual brands. This allowed us to perform a detailed statistical evaluation for each brand within the dataset. Our analysis revealed a significant variance in recall rates among the brands. While some brands exhibited high recall rates, indicating a strong detection capability by Dynaphish, others demonstrated notably low recall rates. 

\begin{figure}[H]
\caption{Statistics of detected phishing samples for each brand in the dataset}
\centering
\includegraphics[width=0.9\textwidth]{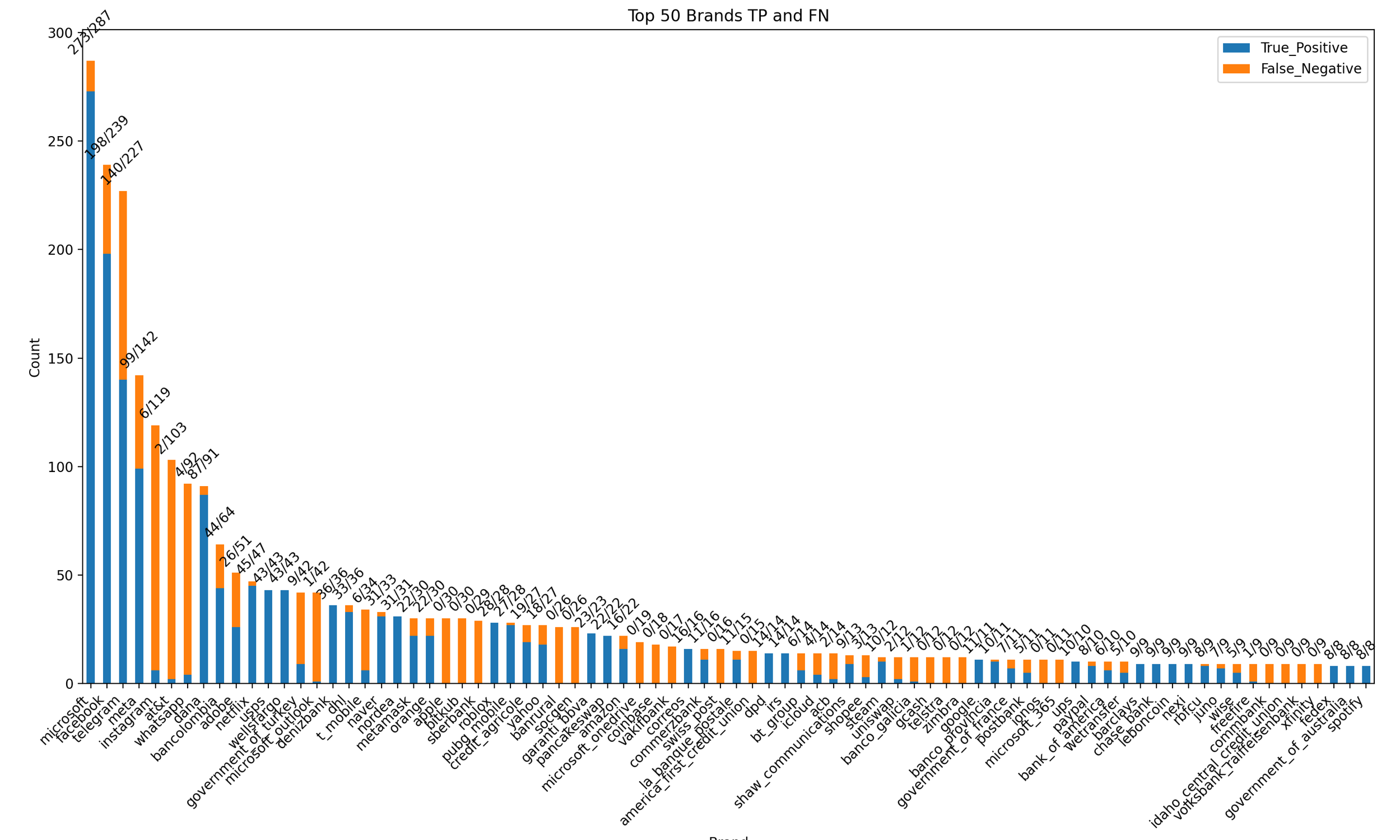}
\end{figure}

To understand the factors that cause such disparity, we dive into the result of brands which has low recall rates and check what has been the factor to impact the detection. 

\subsubsection{Logo Not Detected / Wrong Logo Detected}

Among 3,310 samples, there are 87 samples which logo has not been detected and therefore classified as benign. 

\begin{figure}[H]
\caption{Statistics of false negative samples with no logo detected}
\centering
\includegraphics[width=0.9\textwidth]{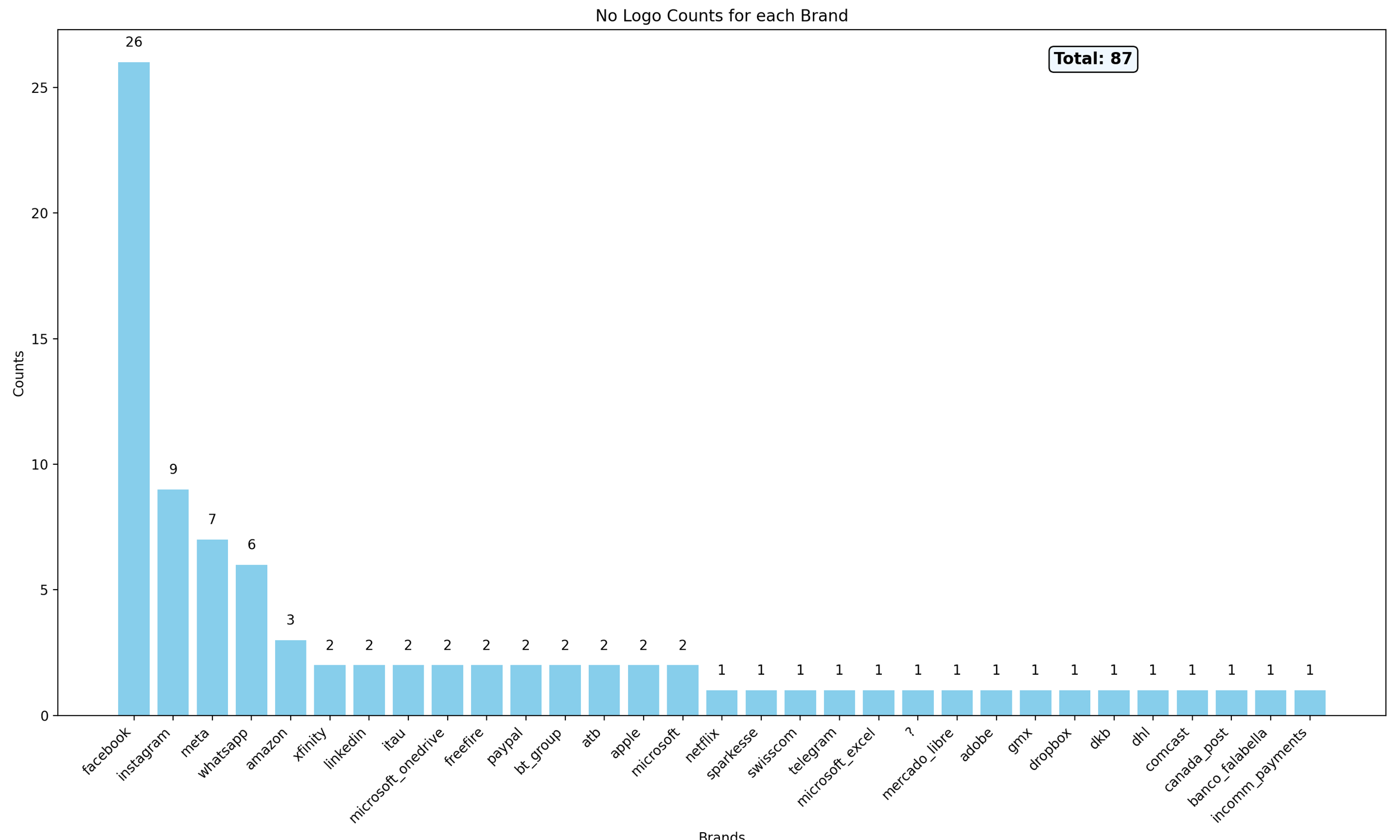}
\end{figure}

This is mainly because these samples do not have logos, or the position of logos are tricky to be detected. This could also caused wrong logo detected, and further impact the result of detection.

\begin{figure}[H]
\caption{The screenshot of phishing sample with no logo detected}
\centering
\includegraphics[width=1\textwidth]{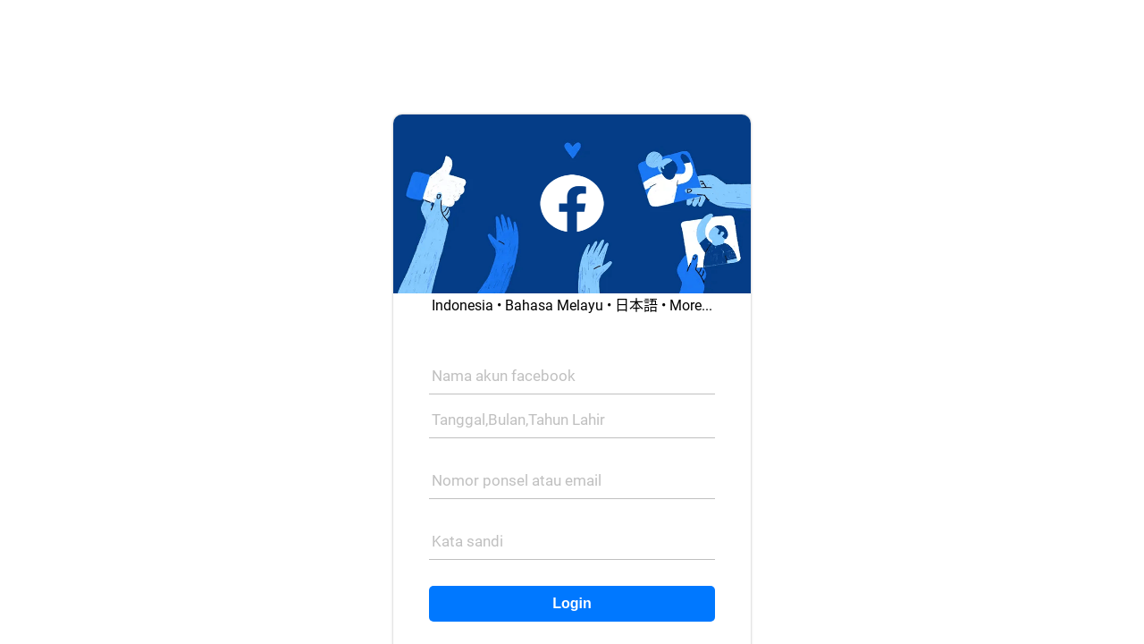}
\end{figure}

\begin{figure}[H]
\caption{The screenshot of phishing sample with wrong logo detected, which lead to wrong result.}
\centering
\includegraphics[width=0.8\textwidth]{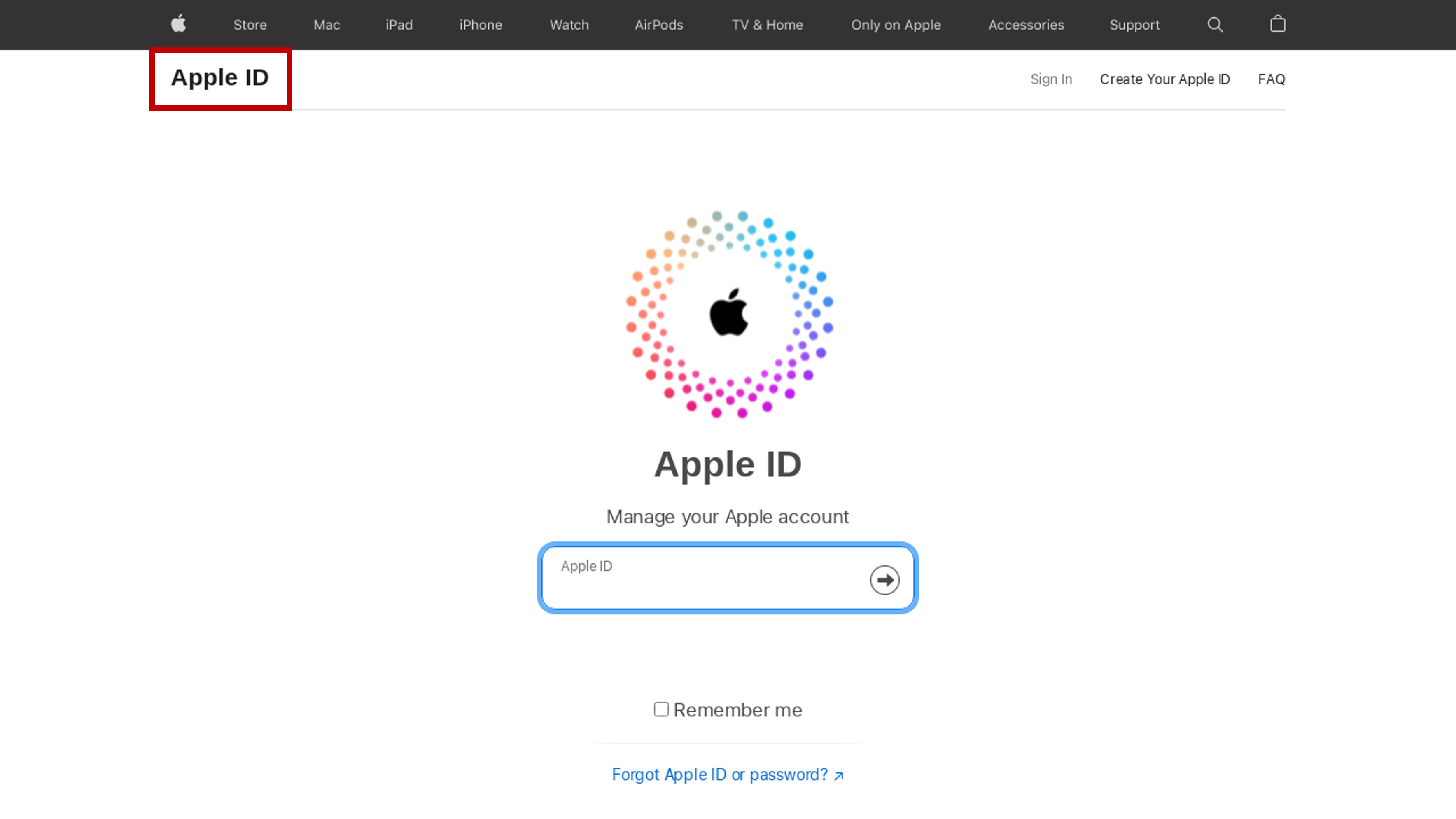}
\label{wrong_logo}
\end{figure}

\subsubsection{Representation Failure}

A critical factor contributing to the elevated false negative rate observed in our analysis is the design of the knowledge expansion module within Dynaphish. The system incorporates a Representation Validation process during the expansion of brand knowledge. Initially, the logo detector isolates the logo from a phishing sample's screenshot. Subsequently, Google's logo detection service is employed to ascertain the brand associated with the logo. 


\begin{figure}[H]
\caption{Dynaphish Representation Validation Overview \cite{liu2023knowledge}}
\centering
\includegraphics[width=0.8\textwidth]{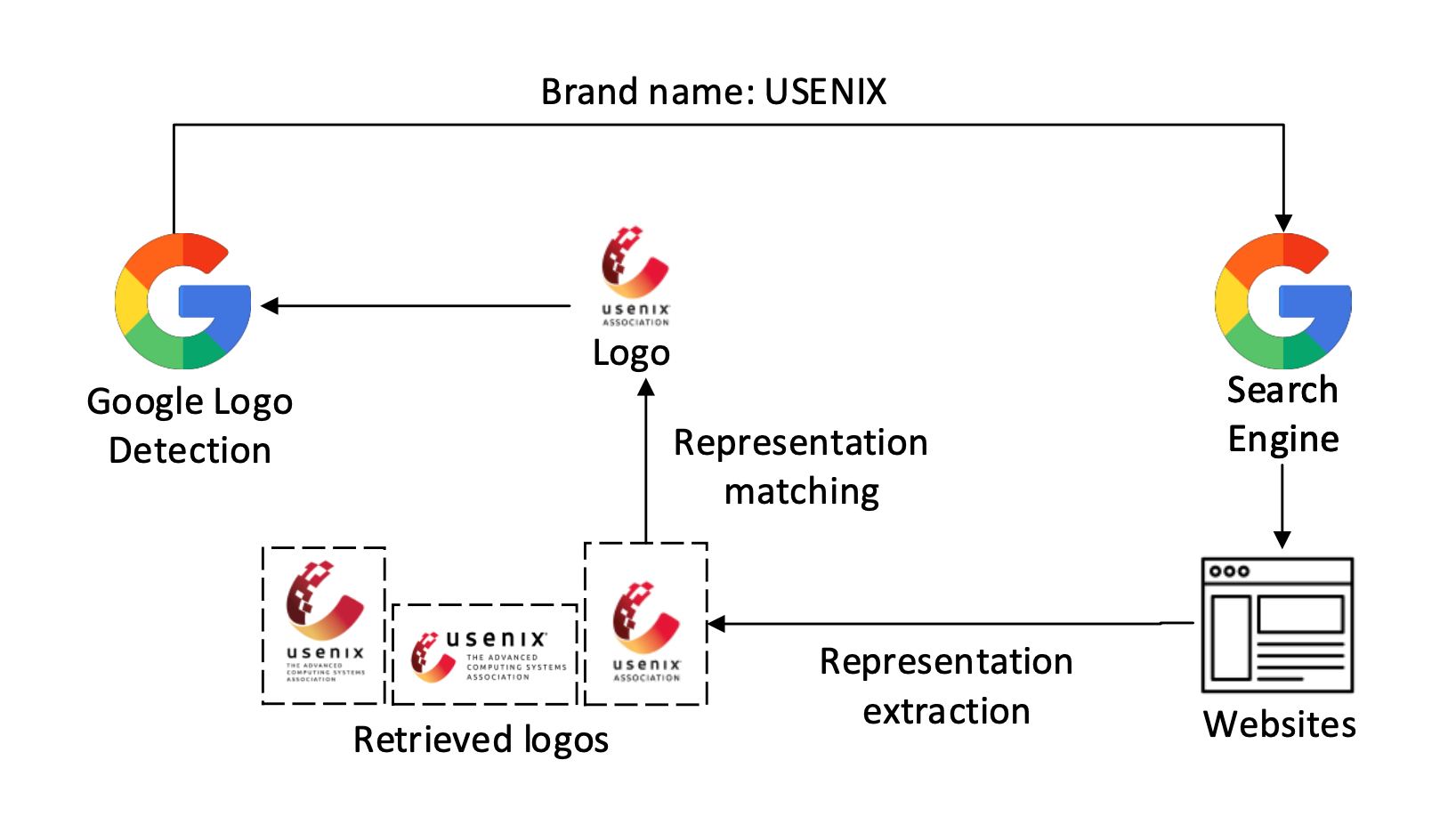}
\end{figure}

The identified brand name is then formulated into a search query, which is input into Google Search. Dynaphish proceeds to extract the top five search results and subjects them to a filtering process. For each link in the refined results list, Dynaphish's driver retrieves the logo and conducts a Representation Validation. If a logo passes this validation with a score that meets the predetermined threshold, the corresponding brand name is incorporated into the target list, thereby broadening the knowledge base. If the sample failed in Representation Validation, this sample will be classified as benign, and the knowledge base will not expand. It is clear that the knowledge base expansion is highly associated with the result of Representation Validation. 

Nonetheless, the knowledge expansion mechanism of Dynaphish is not without its shortcomings. A pivotal issue arises from the reliance on Google logo detector, which can occasionally yield incorrect results. The entire Representation Validation is highly depended on Google Logo detector result, and such inaccuracies inherently affect the subsequent stages of the process, particularly the representation validation. 

Moreover, the reference logos obtained from Google search results are somewhat limited in scope. This limitation becomes apparent when dealing with a variety of logo iterations, as such mechanism does not adequately accommodate the nuances of logo variants. Consequently, this can lead to an incomplete or biased brand representation in the knowledge base. This problem was particularly evident in the detection of AT\&T phishing samples. Out of 103 samples, only 2 were correctly classified as phishing. The samples employed various AT\&T logo variants, yet Dynaphish was only able to retrieve the official logo from AT\&T's website. Consequently, samples featuring alternative logo variants were incorrectly classified as benign. This misclassification was due to the failure to meet the representation validation score threshold, despite the Google logo detector correctly identifying the brand.

\begin{figure}[H]
\caption{AT\&T Official Website Screenshot from Dynaphish Running Records}
\centering
\includegraphics[width=1\textwidth]{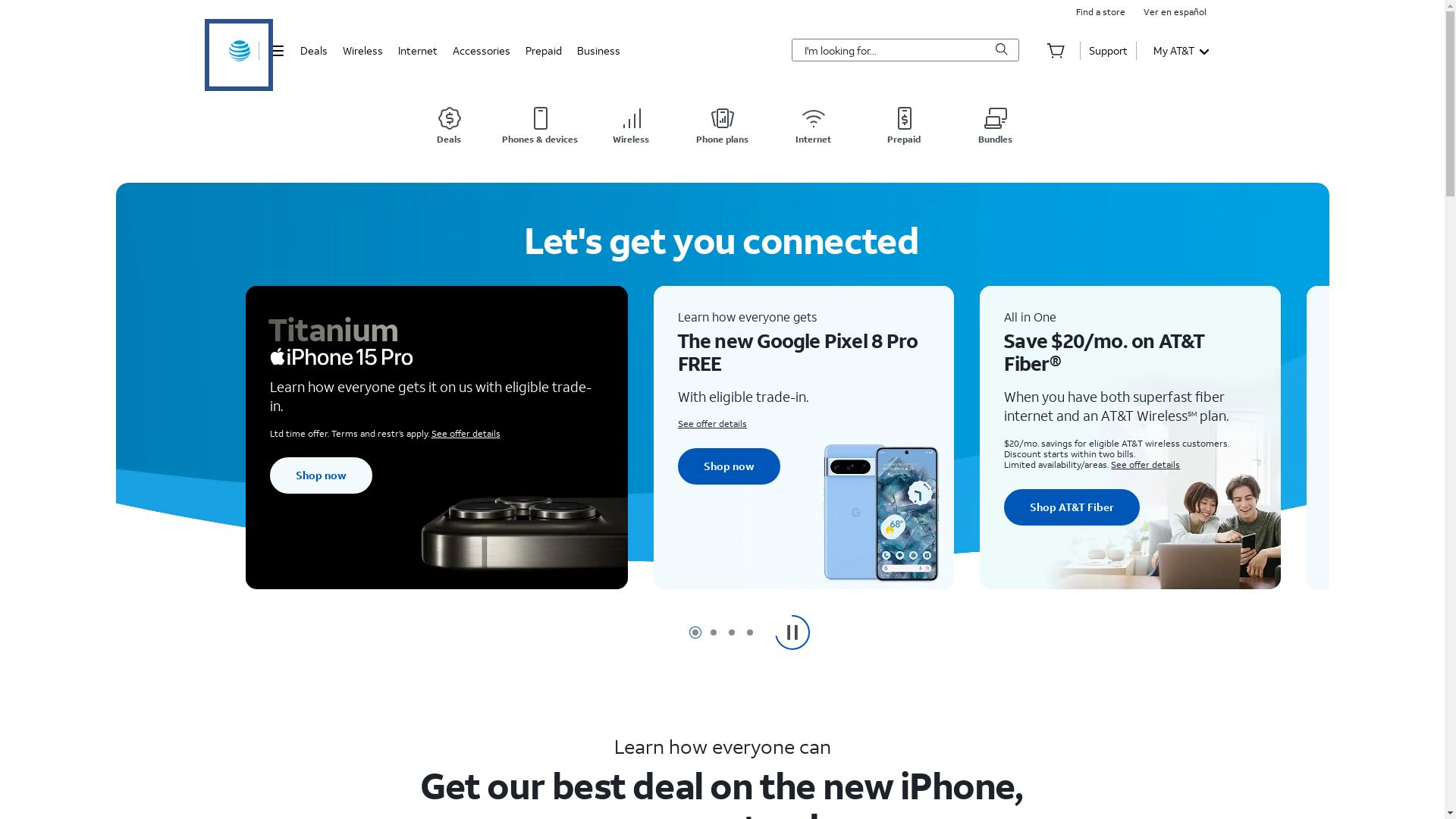}
\end{figure}

\begin{figure}[H]
\caption{AT\&T Phishing Sample which Failed Representation Validation}
\centering
\includegraphics[width=1\textwidth]{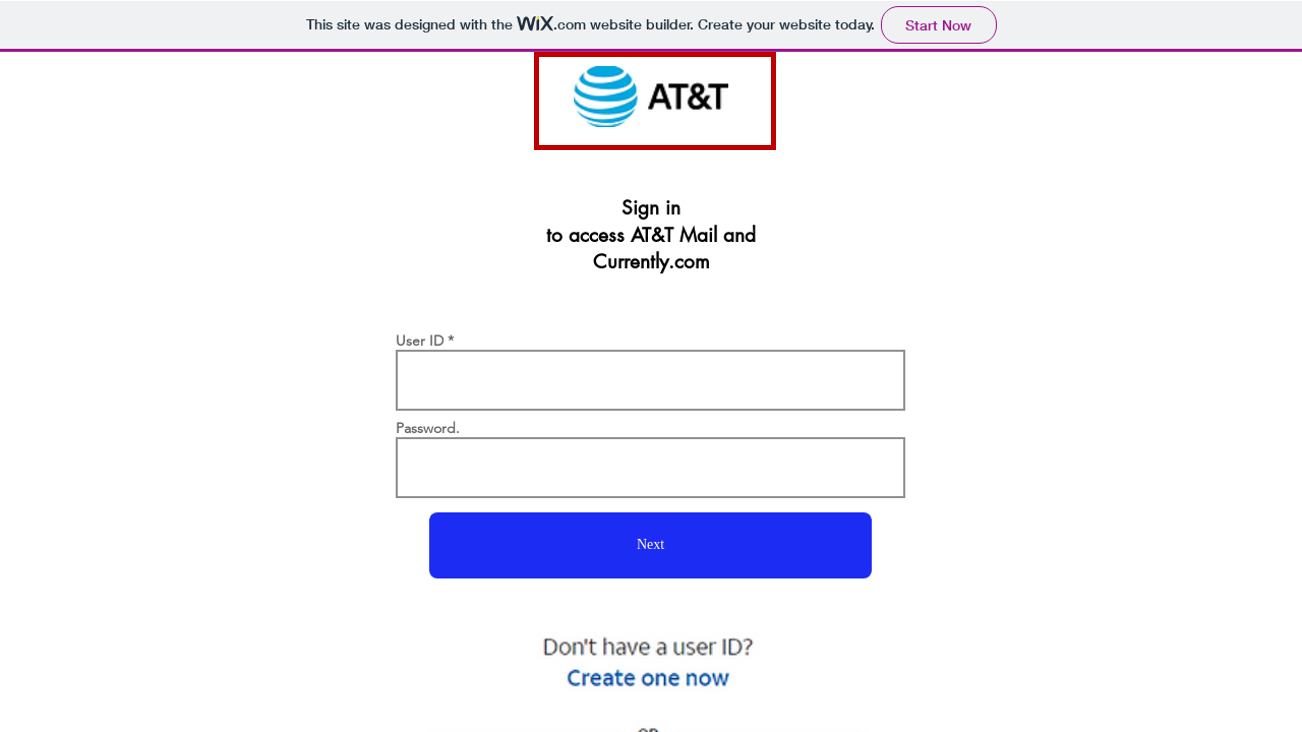}
\end{figure}

Additionally, the filters applied to refine the Google search results can inadvertently impede the representation validation process. If these filters are too restrictive or not aligned with the diverse nature of legitimate brand representations, they may exclude valid logos from the validation process. This exclusion can contribute to an increased rate of false negatives, as legitimate brands may be mistakenly disregarded and not added to the target list, thereby narrowing the ability of knowledge base expansion. A case in point is the detection of Instagram phishing samples. Out of 119 samples, a mere six were identified as phishing. Although the correct logos were cropped and Google's logo search returned accurate results, the domains containing `instagram.com' were automatically excluded due to their presence on the forbidden domain list. As a result, Dynaphish failed to retrieve the correct reference logos from the search results to perform representation score matching which led to a breakdown in both detection and knowledge expansion.

In certain edge cases, Dynaphish encountered difficulties when attempting to retrieve logos from the filtered Google search results. This was particularly noticeable during the identification process for Bitkub phishing samples. Despite the correct logo being cropped and both the Google logo detector and search yielding accurate results, Dynaphish's driver faced challenges with verification pages. This obstacle prevented the successful retrieval of reference logos. Consequently, this led to a complete failure in detecting the Bitkub phishing samples.

\begin{figure}[H]
\caption{Dynaphish Running Screenshot Records when Accessing Google Search Results}
\centering
\includegraphics[width=1\textwidth]{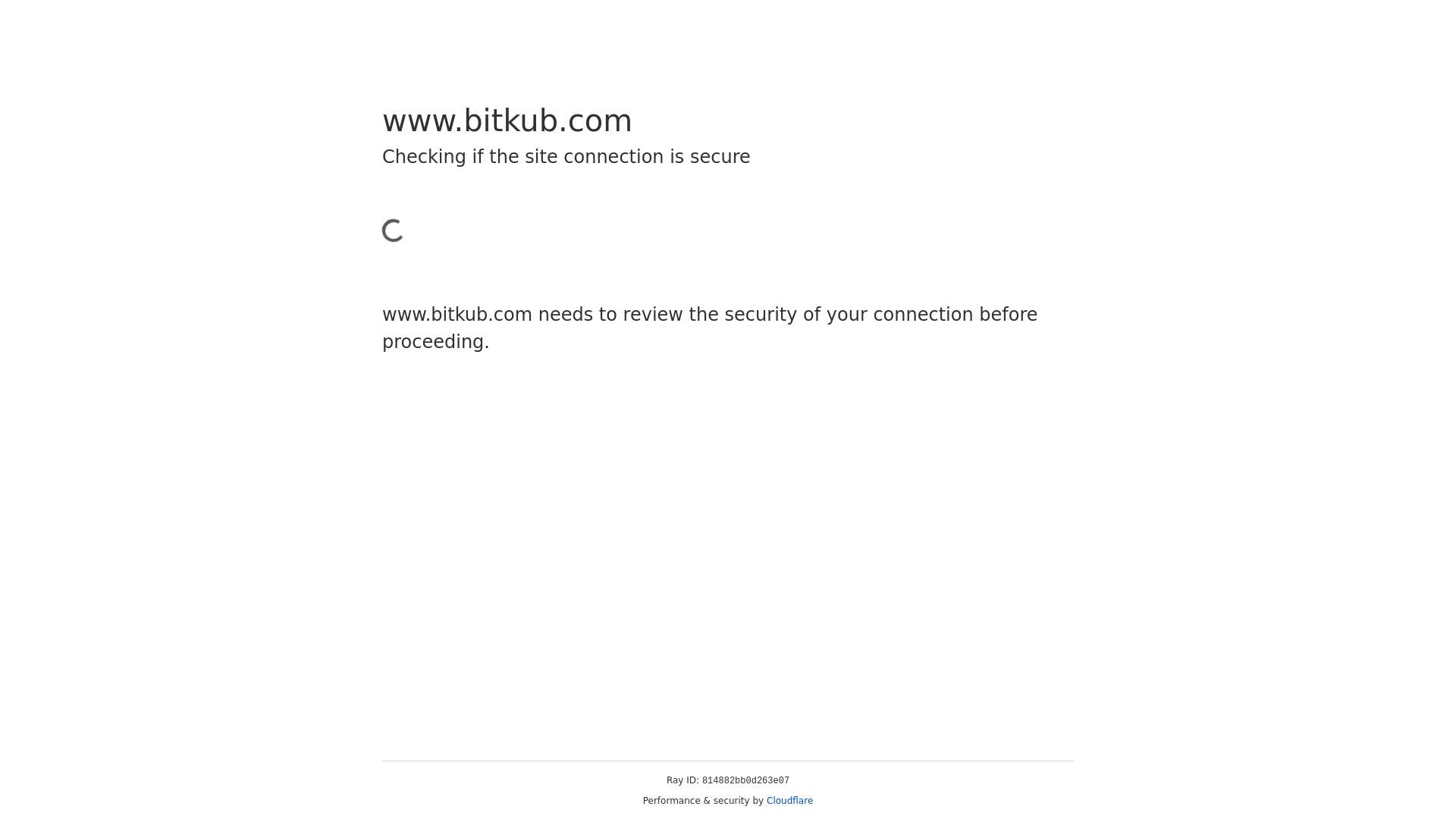}
\end{figure}

\section{Design of Generative AI Enhanced Phishing Agent (GEPAgent)}

Initial experiments with Dynaphish have demonstrated that its capability for information retrieval and processing falls short without a thorough and clean knowledge base. This challenge is amplified in today's rapidly evolving landscape, where new brands emerge daily, making it increasingly difficult to maintain an up-to-date and comprehensive database. To tackle the complexities of reference problem within Dynaphish, we introduce a cutting-edge strategy that capitalizes on the strengths of Large Language Models (LLMs), drawing upon the concept of autonomous agents with the inspiration from the human ability to identify and comprehend phishing webpages. \textbf{G}enerative AI \textbf{E}nhanced \textbf{P}hishing Agent (GEPAgent) is designed to mimic the human approach to collecting and understanding information from webpages. This includes analyzing the webpage's user interface, reading and interpreting the text displayed, and examining the webpage's URL. Such process aims to replicate the nuanced way humans interact with and derive insights from web content. 

In light of the advancements made by OpenAI \cite{openai2023gpt}, with both gpt-3.5-turbo and gpt-4-turbo offering the ability to execute function calls, we recognize the potential of ChatGPT as a powerful agent for information retrieval. These sophisticated models excel at recognizing and interpreting intricate data patterns, thereby markedly improving the precision of brand logo detection and validation processes. ChatGPT's remarkable proficiency in handling text and extracting information makes it an ideal candidate to address the core challenges faced by Dynaphish and other similar reference base solution and also align with our objective of developing an autonomous agent. By incorporating an LLM into our phishing detection framework, our goal is to have better system's proficiency in identifying and verifying a wider array of brand representations. This integration is designed to forge a phishing detection solution of enhanced effectiveness and resilience, by incorporating agents capable of utilizing a variety of tools. Key among these tools are preprocessing utilities and those selectively employed by the agent to refine its analysis and decision-making process. For instance, the Google logo detector allows the agent to identify logos within content for it to better process given information, and by using Google Search as well as Google Image Search, agent could better understand the webpage and extend its knowledge towards the brand, allowing it to make decision with valid reasons.


\begin{figure}[H]
\caption{Overview of Proposed Solution}
\centering
\includegraphics[width=1\textwidth]{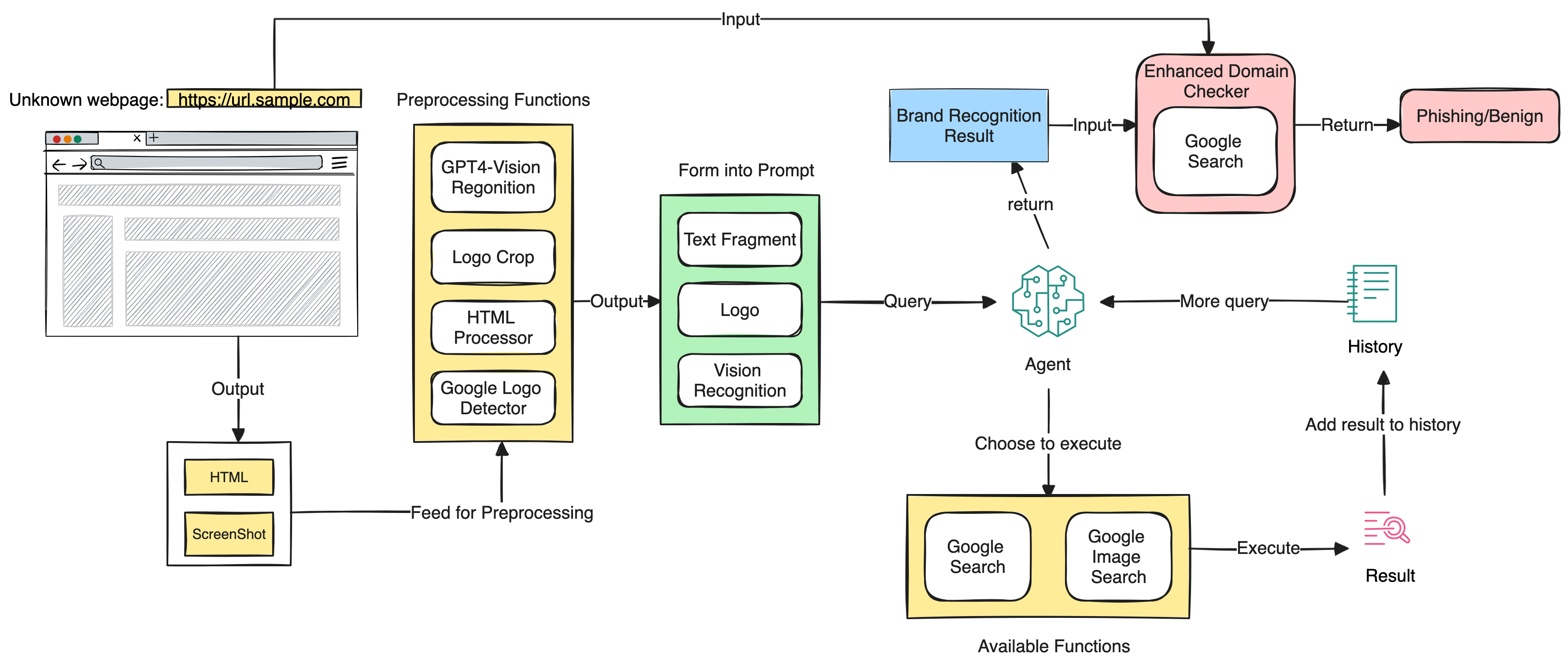}
\end{figure}


\subsection{Information Preprocessing}

To enhance the agent's comprehension of webpages, it's essential to preprocess the various components that users interact with: the URL, text, images, and the rendered user interface. For this purpose, HTML and screenshots are extracted along with the URL, serving as the basis for preprocessing.

The raw HTML is refined to ensure compatibility with GPT's token limit, retaining only crucial elements like the title, input boxes, buttons, and frontend-rendered text. Logo extraction is performed on the screenshot utilizing a logo cropping feature, with the Google logo detector aiding in identifying the webpage's target brand. 

Furthermore, both the logo and the entire screenshot are analyzed by GPT4-V\cite{openai2023gpt}, leveraging its advanced Optical Character Recognition (OCR) and reasoning capabilities to recognize and understand the target brand depicted in the visual content. This step is vital for enabling the agent to accurately interpret and interact with the webpage's information, mirroring the human process of gathering and processing data from digital platforms.

\subsection{Generative Agent}

Utilizing GPT\cite{openai2023gpt}, we simulate an agent that employs its support for function calling and conversation to replicate human-like information gathering and decision-making processes. This agent is designed to acquire additional information using the provided data and toolkits, then make decisions based on the insights it collects. This approach leverages GPT's advanced capabilities to closely mimic the nuanced way humans assess and act upon information, aiming to achieve a higher level of autonomy and intelligence in processing and analyzing web content.  

\subsubsection{Prompt}

We have our customized prompt to settle all the information from previous preprocessing functions:

\lstset{language=Python}
\begin{lstlisting}
prompt = (
        "You are an expert assistant with strong reasoning and brand understanding skills. You're able to identify the brands regardless they are commercial, famous, valuable or not. Be truthful and only make claims that are grounded with the provided information. Do not fabricate any fact or make up non-existent content."
        "Given some text fragments and logo images from a webpage, your final goal is to determine the brand from the Google Search and Google Image Search that matches this webpage, or report that no known brand can be found. Please provide the brand name that matches the given text fragments and logo images, or type 'no brand found' if no brand can be found, and provide reasons for every decision you make. In some cases, if the given tools do not provide enough information, you can change the query to get more information. Try to identify which parts of the input are most likely to be brand-related."
        "You can call any available functions, but only five times in total, and once you make decision, you can give output of brand name with reason in json format and terminate. "
        "INPUT: you will be given: Text fragments from the webpage html, possibly brand-related: "
        f"processed_html: {processed_html}, "
        f"Any possible logo image from the webpage: "
        f"logo: {query_logo}, screen_shot: {screenshot}, Output of Google Logo Search "
        f"on the logo image (e.g., a brand name): {gsearch_detect_res}, this is an important information to identify."
        f"Output of ChatGPT-4 Vision on the screenshot and logo: {gpt4v_res}, the brand of the website as the result is reliable for more than 60 percent of the cases. "
        "TOOLS: To search for information about the text fragments and logo images, select from the following tools to query: Google Search (get_google_search_results). Input: text query; Output: search results, you can use this function to gather more information of text fragment as well as google logo detector result. Google Image Search (get_google_img_search_res): Input: text query; Output: image URLs, basic info, snippets, titles and whether they visually match the current webpage's logo image with similarity score. A score that greater than 0.8 is considered to be similar. You can use it to understand the logo images from the webpage. "
        "FINAL OUTPUT FROM YOU: if you decide to stop, you should only have your output in JSON format for your decision based on your evaluation. There should have two keys: brand_name and reason. Find as much information as possible and specify your decision in 'reason'."
        "If you cannot decide for the brand name, specify 'no brand name' in 'brand_name' and state why you cannot decide in 'reason'. No markdown and indention."
        "Some samples of final output from you is provided below:"
        "SAMPLE FINAL OUTPUT FROM YOU: {\"brand_name\": \"Nike\", \"reason\": \"Based on the google search result and google image result, the logo from the webpage is highly similar to Nike's logo.\"}"
    )
\end{lstlisting}

This initial prompt directs the agent to approach its tasks with integrity, assigning roles that match its capabilities. It offers the essential descriptions needed for processing information and guides the agent on utilizing available toolkits effectively based on its decision. Furthermore, it outlines the specific format and structure the agent's output should adhere to, ensuring that the responses are both relevant and organized according to predefined criteria. This structured approach helps in harnessing the agent's strengths while clearly communicating expectations for its performance and output.

\subsubsection{Function Calls}

According to \cite{openai2023gpt}, users can specify functions, prompting the model to intelligently generate a JSON object filled with arguments needed to call one or several functions. While the Chat Completions API itself doesn't execute these functions, it crafts JSON outputs that, when executed, trigger the desired functions.

For our agent, we have predefined two functions: Google Search and Google Image Search. The Google Logo Detector is integrated during the preprocessing stage to enhance quality of the given information; thus, it's not defined as part of the toolkit in this context.

The agent employs the Google Search function in a manner akin to how humans search for information. It constructs queries based on available data, such as the webpage's title or outcomes from the Google Logo Detector, and other historical data. The search results, including URLs, snippets, and titles, are then compiled with the existing prompt history to form a new, enriched prompt for the agent.

The Google Image Search function operates similarly but includes an additional parameter: a similarity score. This feature is particularly useful for verifying a brand's logo by conducting an image search for the brand's logo. The search returns a list of images, including their sources, thumbnail links, snippets, titles, and contexts. Using the thumbnail link, images are downloaded, and a similarity check is conducted with the cropped logo using a pretrained CV model. The outcome, enriched with this additional verification information, further aids the agent in accurately identifying and validating the target brand of the webpage. The aggregated results are then added to the history, crafting a new prompt that informs subsequent agent inputs.

Equipped with these toolkits, the agent is positioned to significantly enhance its understanding of the brand and the specific webpage in question, thereby advancing its proficiency in brand recognition and identification. Upon reaching a conclusive determination, the agent is designed to issue a stop signal, accompanied by its prediction regarding the target brand and the rationale behind its decision. This process not only streamlines the identification task but also adds a layer of interpretability to the agent's operational logic, making its decisions more transparent and understandable.

\subsection{Domain Checker}

Utilizing the output from the agent, we can conduct a domain check using Google Search. The identified target brand serves as the query for Google Search, from which the top results are compiled into a list with their domain names. If the webpage's URL is found to match any domain in this list, the webpage is deemed benign.

This strategy effectively tackles the issues linked to the knowledge base expansion in Dynaphish by removing the dependency on maintaining a correct and updated knowledge base. Consequently, there's no necessity to verify the accuracy of Google Search results, which in turn eliminates the requirement for result filtering—a process that has been known to introduce a considerable number of false negatives in Dynaphish. The criterion is straightforward: if the webpage's domain aligns with any benign domain identified through Google Search, the webpage is classified as safe. This approach streamlines the verification process and enhances the system's efficiency in distinguishing between benign and phishing sites.




\section{Implementation and Experiment}

To verify the performance of our approach, we conduct a series of experiments to evaluate the performance of our framework. 

\subsection{Implementation of GEPAgent}

Before we implement GEPAgent, we first define which tool should be used for preprocessing, and which tool should be provided to agent. 

Webpage content can be categorized into two types: explicit and implicit information. While both types are inherently static, the knowledge derived from them is dynamic. Explicit information encompasses content directly accessible on the webpage, such as text segments, images, and URLs, which users can immediately perceive. Implicit information, conversely, pertains to the meanings or connections underlying the explicit content, often requiring external sources or prior experience to be fully understood. Gathering this type of information can sometimes be challenging. Our preprocessing functions are designed to transform explicit into implicit information, enhancing the agent's understanding of this information and also equipping the agent with a foundational knowledge base before it begins dynamically expanding its understanding of the webpage.

From this perspective, we can categorize the tools into different processes. Tools that perform static information translation are utilized in preprocessing, whereas tools that facilitate dynamic knowledge expansion are provided as part of the agent's toolkit. Therefore, the Google Logo Detector and GPT4-V are introduced during preprocessing to interpret the information conveyed by images in textual form. Meanwhile, Google Search and Google Image Search are included in the agent's toolkit, enabling it to actively seek out and integrate new information, thereby enhancing its analytical capabilities regarding the webpage.

\subsection{Implementation of Domain Checker}

The domain checker addresses several key issues: selecting the appropriate query to identify the domain of the target brand and executing the actual domain matching. Our strategy employs a domain matching mechanism that includes both top-level and second-level domain matching to effectively tackle the challenge of domain variants. We input the raw brand name enclosed in quotes to generate a list for domain matching. This method anticipates Google Search will yield the webpages associated with the target brand, from which we extract the display link to compile the domain matching list.

Incorporating the brand name and keywords such as ``domain'' or ``official website'' in the search query may lead to receiving irrelevant results from Google Search. For instance, searching for a domain might return information from a WHOIS service that describes the domain registration details of the target brand, which diverges from our intended outcome. To mitigate these issues, our approach focuses on extracting and utilizing the direct display links from search results, ensuring relevance and accuracy in identifying the official domain of the brand.

\subsection{Experiments}

To gain a comprehensive understanding of the capabilities of both GPT-3.5 and GPT-4 in simulating an agent, reasoning, and brand recognition, we structured our evaluation into two distinct segments: Brand Recognition and Phishing Classification. This bifurcated approach allows for a focused assessment of each model's strengths and limitations in these critical areas. 

For the Brand Recognition segment, we aim to determine how effectively each model can identify and differentiate between various brands, an essential skill for accurately assessing the legitimacy of web content. In the Phishing Classification segment, our focus shifts to the models' capacity for discerning between phishing and benign websites. Both experiments leverages the labeled OpenPhish 5k dataset as a source of phishing examples and the Tranco 5k dataset for benign samples.\cite{li2024knowphish}

\subsubsection{Brand Recognition}

Following the implementation of our agent, we embarked on an experiment to assess its brand recognition capabilities. To this end, we randomly selected 200 samples from the labeled OpenPhish 5k dataset as a test set to evaluate the performance of the agent. This approach provides a diverse and challenging set of samples for a comprehensive assessment of the agent's ability to accurately identify and classify brands, allowing us to gauge its effectiveness in real-world phishing detection scenarios.

\begin{table}[h!]
    \centering
    \begin{tabular}{cccc}
     \textbf{Detector} & \textbf{Correct} & \textbf{Wrong} & \textbf{Unknown} \\
     \midrule
      \textbf{Agent-gpt-3.5-turbo}   & 182/200 & 5/200 & 13/200 \\
      \textbf{Agent-gpt-4-turbo}   & 190/200 & 4/200 & 6/200 \\
    \end{tabular}
    \caption{Brand Recognition Result of Agent}
    \label{tab:my_label}
\end{table}

The experimental results highlight the advantages of transitioning from GPT-3.5-turbo to GPT-4-turbo, underscoring the significance of enhanced reasoning capabilities inherent in the latter. Notably, the improvement in brand recognition can be attributed to GPT-4-turbo's advanced reasoning abilities. Additionally, the datasets used to train both models differ, with GPT-4-turbo having access to a broader knowledge base than its predecessor, GPT-3.5-turbo. This expanded knowledge base is likely another critical factor contributing to the improved brand recognition outcomes observed.

Furthermore, we examined the impact of integrating GPT4V into our prompt to assess its influence on the agent's performance. By employing GPT-3.5-turbo as the agent and comparing the results with and without the inclusion of GPT4V vision input, we aimed to evaluate the efficacy of GPT4V in enhancing the agent's ability to recognize brands accurately. 


The integration of GPT4V also showed improvement over scenarios where it was not used. Our investigation reveals that the accuracy of the Google Logo Detector in identifying the correct brand from a provided logo stands at approximately 70\%, a rate significantly dependent on the effectiveness of the logo cropping performed by the pretrained model. In instances where a logo is unavailable or the cropped logo is inaccurately processed, the Google Logo Detector may fail to offer a reliable prediction. Moreover, there is a chance that the Google Logo Detector could yield incorrect results, both of which could negatively impact the agent's decision-making process. As a result, supplementing the process with both the logo and screenshot inputs for GPT4V presents a viable alternative to address these challenges, leveraging its robust OCR capabilities and extensive knowledge base. However, it's important to note that GPT4V has limitations, particularly in handling sensitive content; it may not provide relevant information in cases where the screenshot offers little to no data (e.g., blank screens or Captcha verification pages). 

\subsubsection{Phishing Classification}

To assess the effectiveness of our phishing detection framework comprehensively, we have the entire framework implemented integrated both the agent and the domain checker components. For this evaluation, we utilized the same set of 200 phishing samples previously mentioned, complemented by an additional 200 samples randomly selected from the Tranco 5k benign dataset. This balanced dataset of 400 total samples, evenly divided between phishing and benign examples, provides a robust basis for testing.

The results reveal distinct strengths between GPT-3.5-turbo and GPT-4-turbo in the phishing detection framework. GPT-3.5-turbo demonstrates superior performance in identifying benign samples, whereas GPT-4-turbo excels in detecting phishing samples. This differential performance can be attributed to the handling of samples where the target brand is not identifiable; such samples tend to be classified as benign. Consequently, the rate of False Negatives is closely tied to instances where no brand is detected, as phishing URLs often fail to match legitimate domains.

\begin{table}[h!]
    \centering
    \begin{tabular}{cc}
     \textbf{Detector} & \textbf{TP}  \\
     \midrule
      \textbf{Agent-gpt-3.5-turbo}   & 187/200 \\
      \textbf{Agent-gpt-4-turbo}   & 194/200  \\
    \end{tabular}
    \caption{Phishing Detection Result of Agent}
    \label{tab:my_label}
\end{table}

However, the domain checker also influences the rate of False Positives. In cases where URLs from benign webpages fail the domain check—potentially due to URL redirection or domain variant issues—the framework mistakenly classifies them as phishing. An illustrative example is the Chromedriver by Emburse; the agent accurately predicts "Emburse" as the brand, but the Chromedriver page's domain differs from the official Emburse webpage domains returned by Google Search, leading to a domain check failure. Given GPT-4-turbo's enhanced brand recognition capabilities, there is a heightened risk of increasing False Positive rates during domain checks. 

Beside this experiment, we also conduct the experiment on the domain checker to mitigate the domain variant issue and url redirection. We add one additional using requests library to check whether the url will be redirected if the url failed in domain checking. We also test how the length of the domain list from google search will impact the phishing classification result.

\begin{table}[h!]
    \centering
    \begin{tabular}{p{2.1cm}p{2.1cm}p{2.1cm}p{2.1cm}p{2.1cm}p{2.1cm}}
     \textbf{Detector} & \textbf{Single Domain Match} &\textbf{Single domain match w redirection check} & \textbf{Domain list match(5) with redirection check} & \textbf{Domain list match(5)} & \textbf{Domain list match(10)}\\
     \midrule
      \textbf{Agent-gpt-3.5-turbo}   & 34/200 & 20/200 & 13/200 & 14/200 & 12/200  \\
      \textbf{Agent-gpt-4-turbo}   & 31/200 & 30/200 & 18/200 & 19/200 & 16/200  \\
    \end{tabular}
    \caption{Phishing Detection Result of Agent}
    \label{tab:my_label}
\end{table}

The results indicate that extending the domain list significantly enhances phishing classification performance, especially in cases involving domain variants. Companies or organizations often host similar content across different webpages, employing variant domains for diverse country/language versions. By broadening the scope of domains retrieved from Google Search, the system effectively addresses such scenarios, thereby reducing the incidence of False Positives.

While incorporating a redirection check is believed to further improve performance, challenges arise with webpages that redirect after a certain period. Such delayed redirections are not effectively handled by simple HTTP request libraries and necessitate the use of more sophisticated tools like WebDriver or Selenium for accurate detection. However, deploying these tools for redirection checking significantly increases the runtime cost, particularly when processing phishing samples.

Considering these factors, our finalized approach opts for including 10 domains in the list without implementing a redirection check. This decision strikes a balance between enhancing phishing detection accuracy and maintaining manageable runtime costs, providing an efficient solution for identifying phishing attempts while minimizing the likelihood of falsely flagging benign domains due to variant or redirection-related issues.

\subsection{Baseline} 

To assess the efficacy of our approach comprehensively, we identified a baseline approach that employs similar methodologies, such as the use of LLMs, reliance on external references, or Google Search, for comparison. Thus, we chose Dynaphish as a baseline. In this setup, Dynaphish \cite{liu2023knowledge} operates with an empty knowledge base and web interaction features disabled, allowing us to explore the advantages of deploying an agent over relying solely on a knowledge base.

Additionally, we also examine our approach alongside KnowPhish \cite{li2024knowphish}, which employs Large Language Models (LLMs) for phishing detection but diverges in its strategy by merging a knowledge base with NLP Oneshot prediction focused on HTML files, instead of using an autonomous agent framework. We establish the NLP Oneshot prediction method as an additional baseline to explore how our agent-centric model, enhanced by external tools and the sophisticated analytical power of LLMs, stacks up against a methodology that relies on direct NLP predictions made by LLM on the HTML content of webpages.

\subsubsection{DynaPhish \cite{liu2023knowledge}}

For Dynaphish, we use phishpedia as the model since we do not treat CRP specially in our own phishing detection framework. The same dataset is used (Openphish 5k and tranco 5k) to evaluate the performace of Dynaphish.

\begin{table}[h!]
  \centering
    \caption{Dynaphish Results}
  \begin{tabular}{p{4cm}p{2.5cm}p{2.5cm}p{2.5cm}p{2.5cm}p{2.5cm}}
    \textbf{Detector} & \textbf{TP} & \textbf{TN} & \textbf{FP} & \textbf{FN} \\ 
    \midrule 
    Dynaphish(phishpedia) & 1808/5000 & 4761/5000 & 239/5000 & 3191/5000 \\ 
    & 0.3616 & 0.9522 & 0.0478 & 0.6382 \\ 
  \end{tabular}

  \label{tab:my_label}
\end{table}

The results indicate that while the precision of the detection system is commendably high, the recall remains notably low. This discrepancy primarily arises in the context of phishing sample analysis, where Dynaphish struggles to accurately identify the target brand of a webpage in the absence of a comprehensive knowledge base. The system's reliance on filtering Google search results for brand prediction frequently results in failures, especially with unfamiliar or new brands. Such outcomes highlight Dynaphish's significant dependency on a meticulously curated knowledge base for effective operation. Its existing methodology falls short when confronted with brands not already documented within its database, underscoring a critical limitation in its adaptability and responsiveness to emerging threats.

\subsubsection{NLP Oneshot Prediction \cite{li2024knowphish}}

For this approach, HTML is processed to comply with the token limit of the LLM. The processed HTML will then be fed to the LLM to determine the target brand and its intent, along with reasons provided. This brand prediction is subsequently utilized in conjunction with the domain checker to classify whether the webpage is a phishing site.

\begin{table}[h!]
    \centering
    \begin{tabular}{cccc}
     \textbf{Detector} & \textbf{Correct} & \textbf{Wrong} & \textbf{Unknown} \\
     \midrule
      \textbf{Agent-gpt-3.5-turbo}   & 160/200 & 13/200 & 27/200 \\
    \end{tabular}
    \caption{Brand Recognition and Phishing Classification Result of NLP Oneshot}
    \label{tab:my_label}
\end{table}

According to the results, the rate of accurate brand prediction is relatively lower than that achieved by our approach. This discrepancy suggests that visual information, which may not be captured in the HTML content, plays a critical role and can significantly enhance the accuracy of phishing detection.

\subsection{Evaluation}

Overall, our agent-based approach marks a notable advancement over existing reference-based solutions. By capitalizing on the strong language processing, understanding, and reasoning capabilities of Large Language Models (LLMs), we can more efficiently harness the information available on webpages, leading to improved predictions in phishing detection. Unlike Dynaphish, our method does not depend on a knowledge base, yet it demonstrates superior performance. Additionally, by accommodating more information in various formats and possessing the capability for knowledge expansion, our approach outperforms the NLP oneshot prediction method as well.

\begin{table}[h!]
  \centering
    \caption{Results of Our Approach}
  \begin{tabular}{ccccc}
    \textbf{Detector} & \textbf{Precision} & \textbf{Recall} & \textbf{Accuracy} & \textbf{F1} \\ 
    \midrule 
    agent-gpt-3.5 & \textbf{0.9397} &0.935 &0.9375 &0.9373 \\ 
    agent-gpt-4 & 0.9238 & \textbf{0.97} & \textbf{0.945} & \textbf{0.9463} \\ 
    DynaPhish & 0.8832 & 0.3616 & 0.4999 & 0.5131 \\

  \end{tabular}

  \label{tab:my_label}
\end{table}

\section{Discussion and Conclusion}

For this project, we conducted a thorough investigation and error analysis of existing reference-based phishing detection methods and proposed a novel reference-based framework for automating phishing detection using LLMs and Google's API. Through a series of experiments utilizing a comprehensive dataset, we evaluated the performance of our approach against existing methods. Our findings demonstrate that our agent-based approach achieves superior performance in brand recognition, a critical aspect of reference-based detection, as well as in overall phishing detection capabilities.

\subsection{Limitation}

One of the limitations of our approach is the significantly high runtime cost, which largely depends on the frequency of toolkit usage by the agent. The average runtime for analyzing a single sample is approximately 20 seconds, a duration considered costly. This issue poses a challenge for scalability and real-time application, indicating a need for optimization to strike a balance between efficiency and effectiveness.

Another issue is the number of rounds the agent can use the toolkit for each prediction, especially when the HTML content is lengthy. Lengthy HTML content can impact the number of rounds the agent can utilize the toolkit. Although we have implemented a mechanism to ensure the agent can start running within the token limitation, and we ask the agent to make predictions within 5 rounds, this constraint may not significantly impact the prediction accuracy for the brand. This is because some information may require additional Google searches or Google image searches to gather.

\subsection{Possible Improvement}

According to the experimental results, the agent approach, which consists of multiple rounds of toolkit usage, shows significant improvement over the NLP one-shot approach, which makes predictions within a single prompt. However, further experiments could be conducted by setting a limited number of prompts to observe the results, or by performing an analysis to determine the average number of prompts after which the agent can make a prediction.

Another improvement would be to integrate with a knowledge base, as this could significantly reduce the overall runtime cost and cost of token. By utilizing our approach, we could allow for the expansion of the knowledge base based solely on benign samples, which would also ensure the correctness of the knowledge base.

\bibliographystyle{IEEEtranN}
\bibliography{refs.bib}

\begin{thebibliography}{17}
\providecommand{\natexlab}[1]{#1}
\providecommand{\url}[1]{#1}
\csname url@samestyle\endcsname
\providecommand{\newblock}{\relax}
\providecommand{\bibinfo}[2]{#2}
\providecommand{\BIBentrySTDinterwordspacing}{\spaceskip=0pt\relax}
\providecommand{\BIBentryALTinterwordstretchfactor}{4}
\providecommand{\BIBentryALTinterwordspacing}{\spaceskip=\fontdimen2\font plus
\BIBentryALTinterwordstretchfactor\fontdimen3\font minus \fontdimen4\font\relax}
\providecommand{\BIBforeignlanguage}[2]{{%
\expandafter\ifx\csname l@#1\endcsname\relax
\typeout{** WARNING: IEEEtranN.bst: No hyphenation pattern has been}%
\typeout{** loaded for the language `#1'. Using the pattern for}%
\typeout{** the default language instead.}%
\else
\language=\csname l@#1\endcsname
\fi
#2}}
\providecommand{\BIBdecl}{\relax}
\BIBdecl

\bibitem[Liu et~al.(2023)Liu, Lin, Zhang, Lee, and Dong]{liu2023knowledge}
R.~Liu, Y.~Lin, Y.~Zhang, P.~H. Lee, and J.~S. Dong, ``Knowledge expansion and counterfactual interaction for $\{$Reference-Based$\}$ phishing detection,'' in \emph{32nd USENIX Security Symposium (USENIX Security 23)}, 2023, pp. 4139--4156.

\bibitem[Li et~al.(2024)Li, Huang, Deng, Lock, Cao, Oo, Hooi, and Lim]{li2024knowphish}
Y.~Li, C.~Huang, S.~Deng, M.~L. Lock, T.~Cao, N.~Oo, B.~Hooi, and H.~W. Lim, ``Knowphish: Large language models meet multimodal knowledge graphs for enhancing reference-based phishing detection,'' \emph{arXiv preprint arXiv:2403.02253}, 2024.

\bibitem[Whittaker et~al.(2010)Whittaker, Ryner, and Nazif]{whittaker2010large}
C.~Whittaker, B.~Ryner, and M.~Nazif, ``Large-scale automatic classification of phishing pages,'' 2010.

\bibitem[Khonji et~al.(2013)Khonji, Iraqi, and Jones]{khonji2013phishing}
M.~Khonji, Y.~Iraqi, and A.~Jones, ``Phishing detection: a literature survey,'' \emph{IEEE Communications Surveys \& Tutorials}, vol.~15, no.~4, pp. 2091--2121, 2013.

\bibitem[Lin et~al.(2021)Lin, Liu, Divakaran, Ng, Chan, Lu, Si, Zhang, and Dong]{lin2021phishpedia}
Y.~Lin, R.~Liu, D.~M. Divakaran, J.~Y. Ng, Q.~Z. Chan, Y.~Lu, Y.~Si, F.~Zhang, and J.~S. Dong, ``Phishpedia: A hybrid deep learning based approach to visually identify phishing webpages,'' in \emph{30th USENIX Security Symposium (USENIX Security 21)}, 2021, pp. 3793--3810.

\bibitem[Liu et~al.(2022)Liu, Lin, Yang, Ng, Divakaran, and Dong]{liu2022inferring}
R.~Liu, Y.~Lin, X.~Yang, S.~H. Ng, D.~M. Divakaran, and J.~S. Dong, ``Inferring phishing intention via webpage appearance and dynamics: A deep vision based approach,'' in \emph{31st USENIX Security Symposium (USENIX Security 22)}, 2022, pp. 1633--1650.

\bibitem[Bell and Komisarczuk(2020)]{bell2020analysis}
S.~Bell and P.~Komisarczuk, ``An analysis of phishing blacklists: Google safe browsing, openphish, and phishtank,'' in \emph{Proceedings of the Australasian Computer Science Week Multiconference}, 2020, pp. 1--11.

\bibitem[Sahingoz et~al.(2019)Sahingoz, Buber, Demir, and Diri]{sahingoz2019machine}
O.~K. Sahingoz, E.~Buber, O.~Demir, and B.~Diri, ``Machine learning based phishing detection from urls,'' \emph{Expert Systems with Applications}, vol. 117, pp. 345--357, 2019.

\bibitem[Baidoo-Anu and Ansah(2023)]{baidoo2023education}
D.~Baidoo-Anu and L.~O. Ansah, ``Education in the era of generative artificial intelligence (ai): Understanding the potential benefits of chatgpt in promoting teaching and learning,'' \emph{Journal of AI}, vol.~7, no.~1, pp. 52--62, 2023.

\bibitem[Bösser(2001)]{BOSSER20011002}
\BIBentryALTinterwordspacing
T.~Bösser, ``Autonomous agents,'' in \emph{International Encyclopedia of the Social \& Behavioral Sciences}, N.~J. Smelser and P.~B. Baltes, Eds.\hskip 1em plus 0.5em minus 0.4em\relax Oxford: Pergamon, 2001, pp. 1002--1006. [Online]. Available: \url{https://www.sciencedirect.com/science/article/pii/B0080430767005349}
\BIBentrySTDinterwordspacing

\bibitem[Park et~al.(2023)Park, O'Brien, Cai, Morris, Liang, and Bernstein]{10.1145/3586183.3606763}
\BIBentryALTinterwordspacing
J.~S. Park, J.~O'Brien, C.~J. Cai, M.~R. Morris, P.~Liang, and M.~S. Bernstein, ``Generative agents: Interactive simulacra of human behavior,'' in \emph{Proceedings of the 36th Annual ACM Symposium on User Interface Software and Technology}, ser. UIST '23.\hskip 1em plus 0.5em minus 0.4em\relax New York, NY, USA: Association for Computing Machinery, 2023. [Online]. Available: \url{https://doi.org/10.1145/3586183.3606763}
\BIBentrySTDinterwordspacing

\bibitem[Shen et~al.(2024)Shen, Song, Tan, Li, Lu, and Zhuang]{shen2024hugginggpt}
Y.~Shen, K.~Song, X.~Tan, D.~Li, W.~Lu, and Y.~Zhuang, ``Hugginggpt: Solving ai tasks with chatgpt and its friends in hugging face,'' \emph{Advances in Neural Information Processing Systems}, vol.~36, 2024.

\bibitem[Yang et~al.(2023)Yang, Liu, Han, Chen, Huang, Fu, and Yu]{yang2023appagent}
Z.~Yang, J.~Liu, Y.~Han, X.~Chen, Z.~Huang, B.~Fu, and G.~Yu, ``Appagent: Multimodal agents as smartphone users,'' \emph{arXiv preprint arXiv:2312.13771}, 2023.

\bibitem[Chen et~al.(2024)Chen, Liu, Wang, Zhang, Liu, Lin, Chen, and Zhao]{chen2024agent}
Z.~Chen, K.~Liu, Q.~Wang, W.~Zhang, J.~Liu, D.~Lin, K.~Chen, and F.~Zhao, ``Agent-flan: Designing data and methods of effective agent tuning for large language models,'' \emph{arXiv preprint arXiv:2403.12881}, 2024.

\bibitem[Zeng et~al.(2023)Zeng, Liu, Lu, Wang, Liu, Dong, and Tang]{zeng2023agenttuning}
A.~Zeng, M.~Liu, R.~Lu, B.~Wang, X.~Liu, Y.~Dong, and J.~Tang, ``Agenttuning: Enabling generalized agent abilities for llms,'' \emph{arXiv preprint arXiv:2310.12823}, 2023.

\bibitem[Yao et~al.(2022)Yao, Zhao, Yu, Du, Shafran, Narasimhan, and Cao]{yao2022react}
S.~Yao, J.~Zhao, D.~Yu, N.~Du, I.~Shafran, K.~Narasimhan, and Y.~Cao, ``React: Synergizing reasoning and acting in language models,'' \emph{arXiv preprint arXiv:2210.03629}, 2022.

\bibitem[OpenAI(2023)]{openai2023gpt}
R.~OpenAI, ``Gpt-4 technical report,'' \emph{arXiv}, pp. 2303--08\,774, 2023.

\end{thebibliography}
\end{document}